\RequirePackage{fix-cm} 
\documentclass[a4paper, twoside, reqno, dvips, 12pt]{amsart}
\usepackage{fixltx2e}   

\usepackage[latin1]{inputenc}
\usepackage[T1]{fontenc}

\usepackage{eucal}
\usepackage{esint}
\usepackage{dsfont}
\usepackage{xspace}
\usepackage{amsgen}
\usepackage{amsthm}
\usepackage{amssymb}
\usepackage{amsmath}
\usepackage{upgreek}
\usepackage{amsfonts}
\usepackage{MnSymbol}
\usepackage{mathrsfs}
\usepackage{mathtools}
\usepackage[nice]{nicefrac}

\usepackage{a4wide}

\usepackage[dvipsnames, table]{xcolor}

\definecolor{oneblue}{rgb}{0.0, 0.0, 0.85}
\definecolor{bluepigment}{rgb}{0.2, 0.2, 0.6}
\definecolor{darkgrey}{rgb}{0.273, 0.281, 0.30}
\definecolor{Lightgray}{rgb}{0.89, 0.89, 0.89}
\definecolor{Lightblue}{RGB}{214, 214, 214}
\definecolor{bckg}{RGB}{20.8, 20.8, 20.8} 
\definecolor{charcoal}{rgb}{0.21, 0.27, 0.31}
\definecolor{darkelectricblue}{rgb}{0.33, 0.41, 0.47}

\usepackage[sort&compress, comma, square, numbers]{natbib}

\usepackage{psfrag}
\usepackage{graphicx}
\usepackage{subfigure}
\usepackage{morefloats}
\usepackage{indentfirst}

\usepackage{acronym}
\usepackage{microtype}
\usepackage[labelsep=period,%
            labelfont={bf,sf,color=bluepigment},%
            justification=raggedright]{caption}

\usepackage[usenames, dvipsnames, pdf]{pstricks}
\usepackage{epsfig}
\usepackage{pst-grad} 
\usepackage{pst-plot} 

\usepackage[colorlinks,
           urlcolor=oneblue,
           linkcolor=oneblue,
           citecolor=NavyBlue,
           bookmarksopen=false,
           pdfpagemode=UseNone,
           pagebackref]{hyperref}

\usepackage[explicit]{titlesec}

\titleformat{\section}
  {\color{NavyBlue}\Large\sffamily\bfseries}
  {}
  {0em}
  {\colorbox{bckg!5}{\parbox{\dimexpr\linewidth-2\fboxsep\relax}{\centering\thesection. #1}}}
  [\vspace*{0.33em}]

\titleformat{name=\section,numberless}
  {\color{NavyBlue}\Large\sffamily\bfseries}
  {}
  {0.0em}
  {\colorbox{bckg!10}{\parbox{\dimexpr\linewidth-2\fboxsep\relax}{\centering#1}}}
  [\vspace*{0.33em}]

\titleformat{\subsection}
  {\color{NavyBlue}\large\sffamily\bfseries}
  {}
  {0.0em}
  {\colorbox{bckg!5}{\parbox{\dimexpr\linewidth-2\fboxsep\relax}{\centering\thesubsection. #1}}}
  [\vspace*{0.33em}]

\titleformat{name=\subsection,numberless}
  {\color{NavyBlue}\Large\sffamily\bfseries}
  {}
  {0em}
  {\colorbox{bckg!10}{\parbox{\dimexpr\linewidth-2\fboxsep\relax}{\centering#1}}}
  [\vspace*{0.33em}]

\titleformat{\subsubsection}
  {\color{bluepigment}\sffamily\normalsize\bfseries}
  {\thesubsubsection}
  {0.5em}
  {#1}
  [\vspace*{0.33em}]

\titleformat{\paragraph}[runin]
  {\color{bluepigment}\sffamily\small\bfseries}
  {}
  {0em}
  {#1}

\titlespacing{\section}{1.0em}{1.5em plus 2pt minus 2pt}%
{1.0em plus 2pt minus 2pt}[0em]
\titlespacing{\subsection}{1.0em}{1.5em plus 2pt minus 2pt}%
{1.0em}[0em]
\titlespacing{\subsubsection}{1.0em}{1.5em plus 2pt minus 2pt}%
{1.0em plus 2pt minus 2pt}[0em]

\usepackage{titletoc}

\contentsmargin{0.5em}
\newlength{\tocsep} 
\titlecontents{section}[\tocsep]
  {\addvspace{10pt}\bfseries\sffamily}
  {\contentslabel[\thecontentslabel]{\tocsep}}
  {}
  {\ \titlerule*[0.75pc]{.}\ \thecontentspage}
  []
\titlecontents{subsection}[\tocsep]
  {\addvspace{8pt}\sffamily}
  {\contentslabel[\thecontentslabel]{\tocsep}}
  {}
  {\ \titlerule*[0.5pc]{.}\ \thecontentspage}
  []
\titlecontents*{subsubsection}[\tocsep]
  {\addvspace{2pt}\footnotesize\sffamily}
  {}
  {}
  {\ \titlerule*[0.35pc]{.}\ \thecontentspage}
  [\\*]

\makeatletter
\def\@setauthors{%
  \begingroup
  \def\thanks{\protect\thanks@warning}%
  \trivlist
  \centering\footnotesize \@topsep30\p@\relax
  \advance\@topsep by -\baselineskip
  \item\relax
  \author@andify\authors
  \def\\{\protect\linebreak}%
  \textsc{\normalsize\textcolor{darkelectricblue}{\authors}}%
  \ifx\@empty\contribs
  \else
    ,\penalty-3 \space \@setcontribs
    \@closetoccontribs
  \fi
  \endtrivlist
  \endgroup
}
\def\@settitle{\begin{center}%
  \baselineskip14\p@\relax
    \bfseries
    \textsc{\Large\textcolor{charcoal}{\@title}}
  \end{center}%
}
\makeatother

\usepackage{enumitem}
\setlist[description]{%
  topsep=30pt,               
  itemsep=5pt,               
  font={\bfseries\sffamily\color{NavyBlue}}, 
}

\usepackage{fancyhdr}
\usepackage{lastpage}

\newcommand*\Title{\textcolor{bluepigment}{On the nonlinear dynamics of traveling waves}}
\newcommand*\Authors{\textcolor{bluepigment}{D.~Mitsotakis, D.~Dutykh, \& J.D.~Carter}}
\newcommand*{\plogo}{\textcolor{gray}{{\texttt{arXiv.org} / \textsc{hal}}}} 

\usepackage{acronym}
\acrodef{Serre}{\textsc{Serre}}
\acrodef{cB}{`classical' Boussinesq}
\acrodef{DSW}{Dispesive Shock Wave}
\acrodef{DSWs}{Dispesive Shock Waves}
\acrodef{FEM}{Finite Element Method}
\acrodef{IBVP}{initial-boundary value problem}
\acrodef{RK}{Runge-Kutta}
\acrodef{ODEs}{ordinary differential equations}

\numberwithin{equation}{section}

\newcommand{\R}{\mathbb{R}}
\newcommand{\N}{\mathbb{N}}

\newcommand{\xa}{{x^{\ast}}} 

\newcommand{\ui}{\mathrm{i}}
\newcommand{\ue}{\mathrm{e}}

\newcommand{\ta}{\tau^{\ast}} 

\renewcommand{\P}{\mathbb{P}}
\newcommand{\eps}{\varepsilon}
\renewcommand{\O}{\mathcal{O}}

\renewcommand{\H}{\mathcal{H}}

\newcommand{\Fex}{F_{\rm exact}} 

\renewcommand{\Re}{\operatorname{Re}}

\newcommand{\cf}{\emph{cf.}~}
\newcommand{\ie}{\emph{i.e.}~}

\newcommand{\sech}{\mathrm{sech}}

\begin{document}

\title[\Title]{On the nonlinear dynamics of the traveling-wave solutions of the \textsc{Serre} system}

\author[D.~Mitsotakis]{Dimitrios Mitsotakis$^*$}
\address{Victoria University of Wellington, School of Mathematics, Statistics and Operations Research, PO Box 600, Wellington 6140, New Zealand}
\email{dmitsot@gmail.com}
\urladdr{http://dmitsot.googlepages.com/}
\thanks{$^*$ Corresponding author}

\author[D.~Dutykh]{Denys Dutykh}
\address{LAMA, UMR 5127 CNRS, Universit\'e Savoie Mont Blanc, Campus Scientifique, 73376 Le Bourget-du-Lac Cedex, France}
\email{Denys.Dutykh@univ-savoie.fr}
\urladdr{http://www.denys-dutykh.com/}

\author[J.~Carter]{John D.~Carter}
\address{Mathematics Department, Seattle University, 901 12th Avenue, Seattle, WA 98122, USA}
\email{carterj1@seattleu.edu}
\urladdr{http://fac-staff.seattleu.edu/carterj1/web/}


\begin{titlepage}
\thispagestyle{empty} 
\noindent
{\Large Dimitrios \textsc{Mitsotakis}}\\
{\it\textcolor{gray}{Victoria University of Wellington, New Zealand}}\\[0.02\textheight]
{\Large Denys \textsc{Dutykh}}\\
{\it\textcolor{gray}{CNRS, Universit\'e Savoie Mont Blanc, France}}\\[0.02\textheight]
{\Large John D. \textsc{Carter}}\\
{\it\textcolor{gray}{Seattle University, Washington, USA}}\\[0.16\textheight]

\colorbox{Lightblue}{
  \parbox[t]{1.0\textwidth}{
    \centering\huge\sc
    \vspace*{0.7cm}
    
    \textcolor{bluepigment}{On the nonlinear dynamics of the traveling-wave solutions of the \textsc{Serre} system}

    \vspace*{0.7cm}
  }
}

\vfill 

\raggedleft     
{\large \plogo} 
\end{titlepage}


\newpage
\maketitle
\thispagestyle{empty}


\begin{abstract}

We numerically study nonlinear phenomena related to the dynamics of traveling wave solutions of the \textsc{Serre} equations including the stability, the persistence, the interactions and the breaking of solitary waves. The numerical method utilizes a high-order finite-element method with smooth, periodic splines in space and explicit \textsc{Runge--Kutta} methods in time. Other forms of solutions such as cnoidal waves and dispersive shock waves are also considered. The differences between solutions of the \textsc{\textsc{Serre}} equations and the \textsc{Euler} equations are also studied.

\bigskip
\noindent \textbf{\keywordsname:} Solitary waves; cnoidal waves; stability; finite element method \\


\end{abstract}

\newpage
\tableofcontents
\thispagestyle{empty}


\newpage
\section{Introduction}

The \textsc{Serre} equations (also known as the Green--Naghdi or Su--Gardner equations) \cite{Serre1953a, Su1969, Green1976} approximate the \textsc{Euler} equations of water wave theory and model the one-dimensional, two-way propagation of long waves. If $a$ denotes a typical amplitude of a wave, $d$ the mean depth of the fluid, and $\lambda$ a typical wavelength, then the \textsc{Serre} equations are characterized by the parameters $\eps \doteq a/d = \O(1)$ and $\sigma \doteq d/\lambda \ll 1$, contrary to the Boussinesq equations which model the propagation of small-amplitude, long waves, \ie $\eps \ll 1$ and $\sigma\ll 1$, when the Stokes number is $S \doteq \eps/\sigma^2 = \O(1)$. The Boussinesq equations are often called weakly nonlinear, weakly dispersive equations while the \textsc{Serre} equations are often called fully-nonlinear shallow-water equations.  In dimensionless and scaled variables, the \textsc{Serre} equations take the form:
\begin{equation}\label{eq:serre}
\begin{array}{l}
\eta_t+u_x+\eps(\eta u)_x=0\ ,  \\
u_t+\eta_x+\eps uu_x-\frac{\sigma^2}{3h}[h^3(u_{xt}+\eps uu_{xx}-\eps (u_x)^2)]_x=0\ , 
\end{array}
\end{equation}
for $x\in\R$, $t>0$, along with the initial conditions
\begin{equation}\label{init}
  \eta(x,0)=\eta_0(x)\ ,\quad\quad u(x,0)=u_0(x)\ .
\end{equation}
Here $\eta = \eta(x,t)$ is the free surface displacement, while 
\begin{equation}\label{E2}
  h\doteq 1 + \eps \eta\ ,
\end{equation}
is the total fluid depth, $u = u(x,t)$ is the depth-averaged
horizontal velocity, and $\eta_0$, $u_0$ are given real 
functions, such that $1 + \eps\eta_0=h_0> 0$ for all $x\in
\R$. In these variables, the location of the horizontal bottom is
given by $y = -1$.  For a review of the derivation and the basic
properties of this system we refer to \cite{Barthelemy2004}.

The \textsc{Euler} equations along with the model system (\ref{eq:serre}) admit
traveling wave solutions, \ie waves that propagate without change in
shape or speed \cite{Lavrentiev1947, Benjamin1990, Bona1997}. Solitary waves form
a special class of traveling wave solutions of these systems. The
other important class of traveling wave solutions is the class of
cnoidal wave solutions which can be thought of as the periodic
generalization of solitary waves.  Many Boussinesq-type equations are
known to possess solitary wave and periodic solutions, but do not
admit nontrivial, closed-form solutions.  In contrast, the \textsc{Serre}
equations admit closed-form solitary and cnoidal (periodic) wave
solutions.  The solitary  wave solutions of the \textsc{Serre} system traveling
with constant speed $c_s$ are given by
\begin{equation} \label{eq:solwave}
  h_s(\xi) =(a_0+a_1{\sech}^2(K_s\, \xi)) \big/ \sigma, \quad  u_s(\xi) =c_s\left(1-\frac{a_0}{\sigma h_s(\xi)}\right) \Big/ \epsilon\ ,
\end{equation}
where $\xi = x-c_s t$, $K_s = \sqrt{3 a_1/4\sigma a_0^2 c_s^2}$, $c_s = \sqrt{(a_0 + a_1)/\sigma}$, $a_0 > 0$, and $a_1 > 0$. By taking $a_0 = \sigma$ and $a_1 = \eps \sigma A_s$ the formulas for the classical solitary waves that are
homoclinic to the origin are obtained.

The cnoidal waves of the \textsc{Serre} system traveling with constant
speed $c_c$ are given by  
\begin{equation}
h_c(\xi) = (a_0 + a_1 {\rm dn}^2(K_c\,\xi, k)) \big/ \sigma\ , \quad u_c(\xi) = c_c\left(1 - \frac{h_0}{h(\xi)}\right) \Big/ \epsilon\ , \label{eq:cnwave} 
\end{equation}
where $h_0\ =\ a_0\ +\ a_1E(m)/K(m)$, $K_c\ =\ \sqrt{3a_1}/2\sqrt{a_0(a_0\ +\ a_1)(a_0\ +\ (1-k^2)a_1)}$, $c_c\ =\ \sqrt{a_0(a_0\ +\ a_1)(a_0\ +\ (1\ -\ k^2)a_1)/\sigma h_0^2}$ , 
$k\ \in\ [0,1]$, $m\ =\ k^2$, $a_0\ >\ 0$, and $a_1\ >\ 0$. Here $K$ and $E$
are the complete elliptic integrals of the first and second kind
respectively.  Note that (\ref{eq:solwave}) are
the $k\rightarrow1$ limit of (\ref{eq:cnwave}).

Another fundamental property of the \textsc{Serre} system is the
conservation of the energy which plays also the role of the
Hamiltonian, $ \H(t) = \frac{1}{2}\int_{-\infty}^{\infty} (\eps h u^2+\frac{\eps\sigma^2}{3}h^3u_x^2+\eps\eta^2) d x$, 
in the sense that $\H(t) = \H(0)$ for all $t > 0$ up to the maximal
time $T$ of the existence of the solution. 

In this paper we study the problem of the nonlinear stability (orbital
and asymptotic) of the traveling waves of the \textsc{Serre} system by using
numerical techniques.  We provide numerical evidence of stability with
respect to certain classes of perturbations.  Phenomena such as
perturbations of the traveling waves, perturbations of the \textsc{Serre}
system and interactions of traveling waves are studied analyzing the
stability properties of the waves at hand.  We also study the
interactions of dispersive shock waves (DSWs) in the \textsc{Serre} system.
The physical relevance of the \textsc{Serre} equations is addressed whenever
possible.

The paper is organized as follows. The numerical method is presented
briefly in Section~\ref{sec:numerscheme}. The compatibility of the
solitary waves of the \textsc{Serre} and the \textsc{Euler} systems is examined in
Section~\ref{sec:travelwaves}. The head-on collision of solitary waves
is studied in Section~\ref{sec:head-on}. A number of issues related to
the stability of the traveling waves are discussed in
Sections~\ref{sec:stability}. The interaction of DSWs is presented in
Section~\ref{sec:dswinter}.   


\section{The numerical method}\label{sec:numerscheme}

The numerical method of preference is a high-order
Galerkin / Finite element method (FEM) for the spatial discretization
combined with the classical fourth-order explicit Runge--Kutta method
in time. In some cases adaptive time-stepping methods, such as the
Runge--Kutta--Fehlberg, the Cash--Karp and the Dormand--Prince
methods~\cite{Hairer2009}, were employed to verify that there are no
spurious solutions or blow-up phenomena.  This numerical scheme has
been shown to be highly accurate and stable since there is no need for
a restrictive condition on the step-size but only mild conditions of
the form $\Delta t\leq C\Delta x$ \cf \cite{Mitsotakis2014}.  The
conservation of the Hamiltonian was monitored and was usually
conserved to within 8 to 10 significant digits.  In order
to ensure the accuracy of the numerical results obtained with the FEM
we compared most with the analogous results obtained with the
pseudo-spectral method described and analyzed in
\cite{Dutykh2011a}. The experiments presented in this paper also serve
as numerical benchmarks for the efficacy of the numerical scheme.

We consider (\ref{eq:serre}) with periodic boundary conditions and,
for simplicity, assume $\varepsilon = \sigma = 1$.  We rewrite
(\ref{eq:serre}) in terms of $(h,u)$ rather than $(\eta,u)$. This is
done by using (\ref{E2}) and yields the initial-boundary value problem

\begin{equation}\label{eq:serre2}
\begin{array}{l}
h_t + (hu)_x = 0\ , \\
u_t + h_x + uu_x - \frac{1}{3h}\left[h^3(u_{xt} + uu_{xx} - (u_x)^2\right]_x = 0\ ,  \\
\partial_x^i h(a,t) = \partial_x^i h(b,t), \quad i = 0,1,2,\ldots\ ,  \\
\partial_x^i u(a,t) = \partial_x^i u(b,t), \quad i = 0,1,2,\ldots\ , \\
h(x,0) = h_0(x)\ ,  \\
u(x,0) = u_0(x)\ , 
\end{array} 
\end{equation}
where $x \in [a,\ b]\subset \R$ and $t \in [0,\ T]$. Considering a
spatial grid $x_i = a+i\ \Delta x$, where $i=0,1,\cdots, N$, $\Delta
x$ is the spatial mesh length, and $N\in \N$, such that $\Delta x =
(b-a)/N$.  We define the space of the periodic cubic splines 
\begin{equation*}
  S = \left\{\phi\in C^2_{\mathrm{per}} [a,b]\Big| \phi|_{[x_i, x_{i+1}]} \in \P^{3},\ 0\leq i\leq N-1 \right\}\ ,
\end{equation*}
where $C^2_{\mathrm{per}} = \left\{f\in C^2 [a,b]\Big| f^{(k)}(a) = f^{(k)}(b),\ 0\leq k\leq r\right\}$ and $\P^k$ is the space of polynomials of degree $k$. The semi-discrete scheme is reduced to finding $\tilde{h}$, $\tilde u\in S$ such that
  \begin{equation}
  \begin{array}{l}
(\tilde{h}_t,\phi)+\left((\tilde{h}\tilde{u})_x,\phi \right)=0\ ,  \\
\mathcal{B}(\tilde{u}_t,\phi;\tilde{h})+\left(\tilde{h}(\tilde{h}_x+\tilde{u}\tilde{u}_x),\phi \right)+\frac{1}{3}\left(\tilde{h}^3(\tilde{u}\tilde{u}_{xx} - (\tilde{u}_x)^2),\phi_x)\right) = 0\ , 
  \end{array}\label{eq:semid}
  \end{equation}
where $\mathcal{B}$ is defined as the bilinear form that for fixed
$\tilde{h}$ is given by 
\begin{equation}\label{eq:bilinear}
  \mathcal{B}(\psi,\phi;\tilde{h})\doteq (\tilde{h}\psi,\phi) + \frac{1}{3}(\tilde{h}^3\psi_x,\phi_x)\ \mbox{for } \phi,\psi\in S\ .
\end{equation}
The system of equations (\ref{eq:semid}) is accompanied by the initial conditions 
\begin{equation}\label{eq:initconds}
  \tilde{h}(x,0) = \mathcal{P}\{h_0(x)\}\ ,\quad 
  \tilde{u}(x,0) = \mathcal{P}\{u_0(x) \}\ ,
\end{equation}
where $\mathcal{P}$ is the $L^2$-projection onto $S$ satisfying
$(\mathcal{P}v,\phi) = (v,\phi)$ for all $\phi\in S$. Upon choosing
appropriate basis functions for $S$, (\ref{eq:semid}) is a system of
ODEs. For the integration in time of this system, we employ the
classical, four-stage, fourth-order explicit Runge--Kutta method.

\section{Solitary waves}\label{sec:travelwaves}

In this section we study how close are the solitary waves of the \textsc{Euler} equations to those of the \textsc{Serre} equations. In other words we verify the consistency of the \textsc{Serre} equations and the ability to approximate well the basic solitary wave dynamics of the \textsc{Euler} equations. 

\subsection{Consistency of solitary waves}

The \textsc{Serre} system and the \textsc{Euler} equations both possess solitary waves
that decay exponentially to zero at infinity.  Although the
justification of the \textsc{Serre} equations ensures that its solutions will
remain close to \textsc{Euler} solutions, it is not known how close remain an \textsc{Euler}
solitary wave to a \textsc{Serre} solitary wave when it is used as initial
condition to the \textsc{Serre} system.

While the \textsc{Serre} system admits solitary wave solutions of the form given in (\ref{eq:solwave}), there are no known closed-form solitary wave solutions of the \textsc{Euler} system but only Fenton's asymptotic solution \cite{Fenton1972}. Although this solution is an accurate approximation, modern numerical techniques enable us to compute solitary waves of the \textsc{Euler} equation even more accurately. For this reason we compute \textsc{Euler} solitary waves numerically. The numerical method is a \textsc{Petviashvili} iteration applied to the \textsc{Babenko} equation, \cite{Petviashvili1976, Clamond2012b, Dutykh2013b}. In order to integrate the full \textsc{Euler} equations in time, we employ the method of holomorphic variables. This formulation was first coined by L.~\textsc{Ovsyannikov} (1974) \cite{Ovsyannikov1974} and developed later by A.~\textsc{Dyachenko} \emph{et al}. (1996) \cite{Dyachenko1996} in deep waters. The extension to the finite depth case was given in \cite{Li2004}. The resulting formulation is discretized in the conformal domain using a Fourier-type pseudo-spectral method. For the time integration we employ an embedded Runge--Kutta scheme of 5(4)th order along with the integrating factor technique to treat the dispersive linear part.

In order to demonstrate the ability of the \textsc{Serre} equations to
approximate the \textsc{Euler} equations, we first compare the characteristics
of two solitary waves with speeds $c_s = 1.1$ and $c_s = 1.2$.
The solitary waves are not identical but their
differences are small and more pronounced at the higher speed.  For example, a
$10\%$ increase in the speed leads to an increase in solitary wave
amplitude of almost $50\%$ while the normalized difference between the
\textsc{Euler} and \textsc{Serre} solitary waves increased by more than a factor of
two.  The amplitude of several \textsc{Euler} and \textsc{Serre} solitary waves are
presented in Table~\ref{T6}.

\begin{table}[ht]
\center
  \begin{tabular}{lll}
    \hline
    $c_s$ & \textsc{Euler} & \textsc{Serre} \\
    \hline
    $1.01$ & $0.02012$ & $0.0201$ \\
    $1.05$ & $0.10308$ & $0.1025$ \\
    $1.1$ & $0.21276$ & $0.2100$ \\
    $1.15$ & $0.33007$ & $0.3225$ \\
    $1.2$ & $0.45715$ & $0.4400$ \\
    $1.28$ & $0.70512$ & $0.6384$ \\
    \hline
  \end{tabular}
  \bigskip
  \caption{\small\em Amplitudes of the \textsc{Euler} and \textsc{Serre} solitary waves corresponding to different speeds.}
  \label{T6}
\end{table}

\begin{figure}
  \bigskip
  \centering
  {\includegraphics[width=0.99\columnwidth]{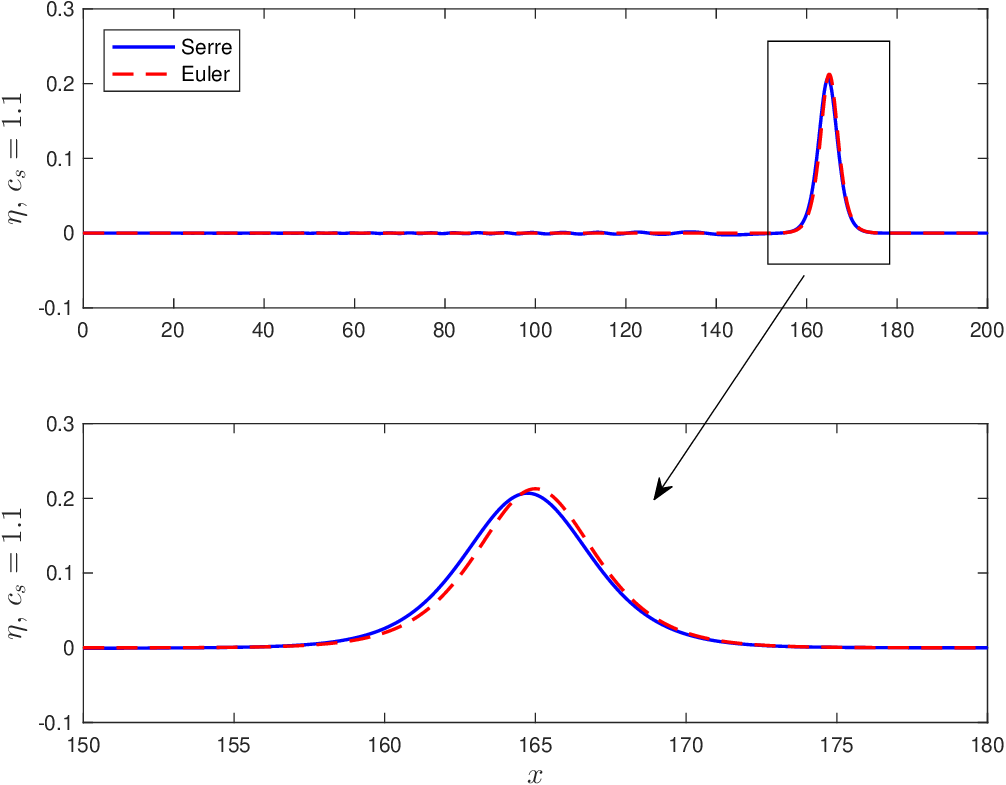}}
  \caption{\small\em The evolution of an \textsc{Euler} solitary wave with $c_s = 1.1$ when used as initial condition in the \textsc{\textsc{Serre}} system.}
\label{fig:figure3}
\end{figure}

\begin{figure}
  \bigskip
  \centering
  {\includegraphics[width=0.99\columnwidth]{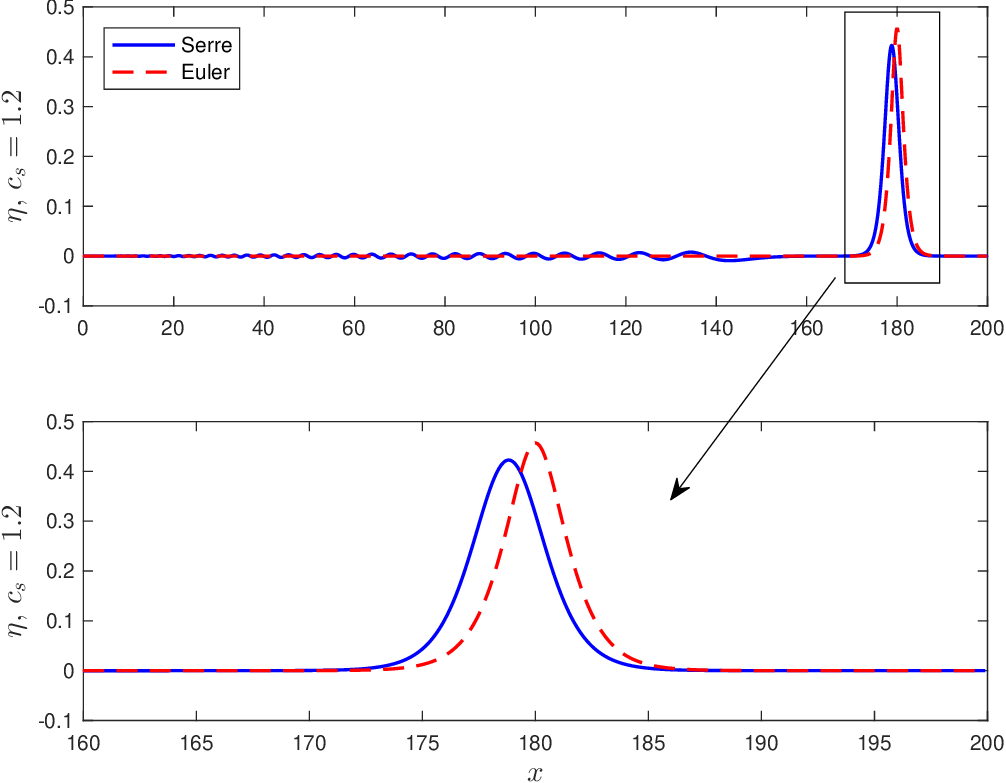}}
  \caption{\small\em The evolution of an \textsc{Euler} solitary wave with $c_s=1.2$ when used as initial condition in the \textsc{Serre} system.}
  \label{fig:figure4}
\end{figure}

Next, we examine how the solitary waves of the \textsc{Euler} system propagate
when they are used as initial conditions to the \textsc{Serre} system.
Specifically, we use the numerically generated solitary wave solutions
of the \textsc{Euler} equations and the exact formula $u=c_s\eta/(1+\eta)$ to
define the initial conditions $\eta_0$ and $u_0$ for the \textsc{\textsc{Serre}}
equations.  Then, we numerically integrate the \textsc{Serre} system.
Figures~\ref{fig:figure3} and \ref{fig:figure4} contain plots of the
solutions at $t = 150$ obtained using the \textsc{Euler} solitary waves with
$c_s = 1.1$ and $c_s = 1.2$.  These figures demonstrate that the
difference between the \textsc{Euler} solitary wave and numerical \textsc{\textsc{Serre}}
solution is greater when $c_s=1.2$ than when $c_s=1.1$.  We note that
the value $c_s = 1.2$ is a relatively large value since the largest
value we can use to generate an \textsc{Euler} solitary wave is
$c_s=1.29421$. 

\begin{figure}
  \bigskip
  \centering
  {\includegraphics[width=0.79\columnwidth]{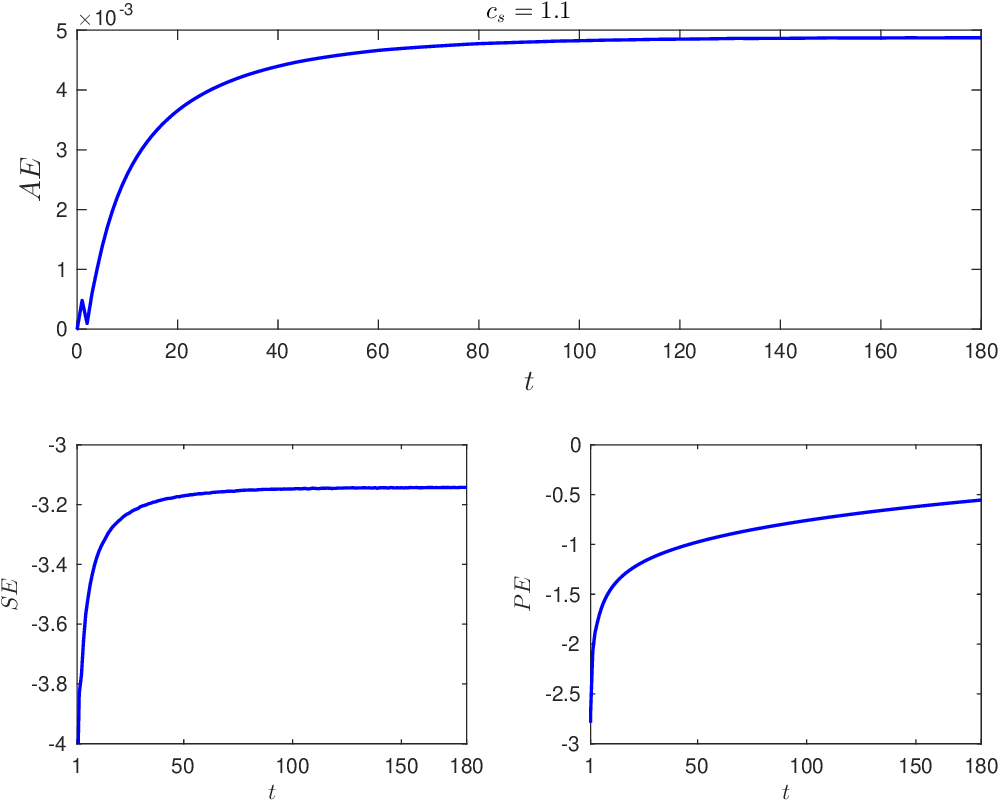}}
  \caption{\small\em The amplitude, shape and phase error of the Euler's solitary wave of $c_s = 1.1$ propagating with \textsc{Serre} equations. See also Figure~\ref{fig:figure3}.}
  \label{fig:figure5}
\end{figure}

\begin{figure}
  \bigskip
  \centering
  {\includegraphics[width=0.79\columnwidth]{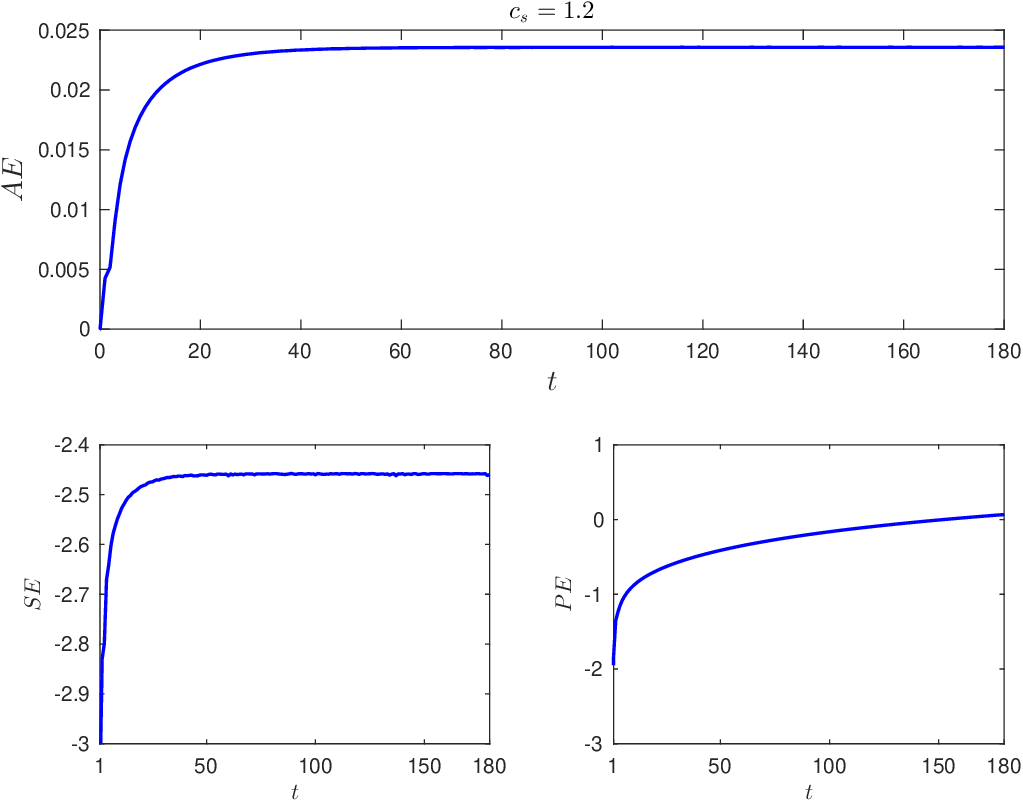}}
  \caption{\small\em The amplitude, shape and phase errors of the Euler's solitary wave of $c_s = 1.1$ propagating with \textsc{\textsc{Serre}} equations. See also Figure~\ref{fig:figure4}.}
  \label{fig:figure6}
\end{figure}

To study further the differences between the \textsc{Euler} and \textsc{Serre} solitary
waves, we consider three quantities pertinent to the propagation of
the solitary waves: the \emph{amplitude, shape} and \emph{phase}. First,
we define the normalized peak amplitude error as 
\begin{equation}\label{eq:AE}
  AE[F] \ \doteq \ \frac{ \left| F(\xa(t),t) - F(0,0)\right|  }{ \left|F(0,0) \right|}~,
\end{equation}
where $\xa(t)$ is the curve along which the computed solution $F(x,t)$
achieves its maximum. Monitoring $AE$ as a function of time, we
observe that although the \textsc{Euler} solitary waves do not propagate as
traveling waves to the \textsc{Serre} system, their amplitude asymptotically
tends towards a constant indicating that they evolve into a solitary
wave solution of the \textsc{Serre} equation, see Figures~\ref{fig:figure5}
and \ref{fig:figure6}.  

We define the normalized shape error as 
\begin{equation}\label{eq:shape}
  SE[F] \ \doteq \  \log_{10}(\min_\tau \zeta(\tau))~, \quad \zeta(\tau) \doteq \ \frac{\|F(x,t^n) - \Fex(x,\tau)\| }{ \|\Fex(x,0)\| }~.
\end{equation}
The minimum in (\ref{eq:shape}) is attained at some critical $\tau = \ta(t^n)$. This, in turn, is used to define the phase error as
\begin{equation}\label{eq:phase}
  PE[F] \ \doteq \  \log_{10}(|\ta - t^n|).
\end{equation}
In order to find $\tau^*$, we use Newton's method to solve the
equation $\zeta''(\tau) = 0$. The initial guess for Newton's method is
chosen as $\tau^0 = t^n - \Delta t$. Figures~\ref{fig:figure5} and
\ref{fig:figure6} contain plots of the shape and phase errors.  We
observe that the shape error is of $\O(10^{-3})$ when $c_s = 1.1$ and
of $\O(10^{-2})$ when $c_s = 1.2$. The phase error increases since the
solitary waves propagate with different speeds. It is remarkable that
the phase speeds of the new solitary waves of the \textsc{Serre} system are
almost the same as the phase speeds of the Euler's solitary waves. For
example, the speeds are $c_s \approx 1.09$ and $c_s \approx 1.19$.
Similar comparisons have been performed for other model equations such
as the classical Boussinesq system and the results are
comparable~\cite{BCS,Duran2013}. 

\subsection{Head-on collision of solitary waves}
\label{sec:head-on}

The collision of two solitary waves of the \textsc{Serre} system has
previously been studied theoretically and numerically
in~\cite{Mitsotakis2014, Dutykh2011a, Li2004, Mirie1982, Su1980, Choi1999, CGHHS, Marchant1990}. While the phenomena related to these interactions have been understood quite well we summarize here the dynamics of the head-on collision of solitary waves and we focus on the related dynamics compared to experimental data and to  numerical simulations of the full \textsc{Euler} equations. The interaction of solitary waves for the \textsc{Serre} equations is general more inelastic than in weakly nonlinear models such as the classical
Boussinesq system~\cite{BCS}. Highly nonlinear interactions result in
the generation of large amplitude dispersive tails. 

\begin{figure}
  \bigskip
  \centering
  {\includegraphics[width=0.99\columnwidth]{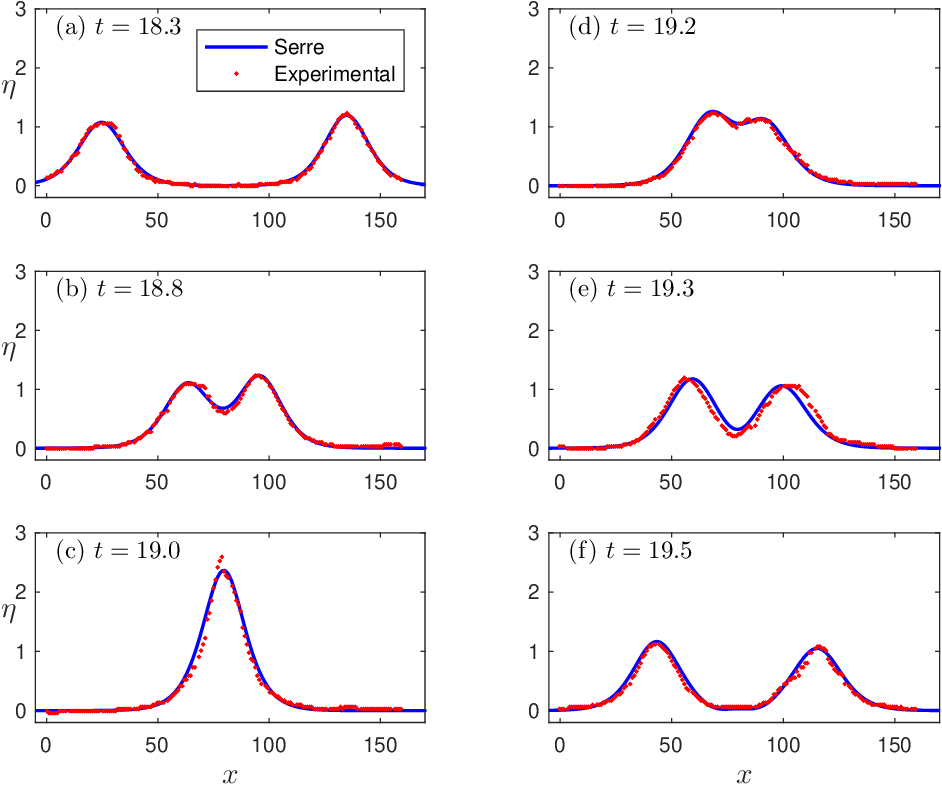}}
  \caption{\small\em Comparison of the head-on collision of two solitary waves of the \textsc{Serre} system with experimental data.}
  \label{fig:figure7}
\end{figure}

In order to study the physical relevance of the head-on collision of
two \textsc{Serre} solitary waves, we compare the \textsc{Serre} numerical solution with
the experimental data of \cite{CGHHS}.  In this experiment, the \textsc{Serre}
system is written in dimensional and unscaled form with an initial
condition that includes two counter-propagating solitary waves in the
interval $[-5, 5]$.  The speeds of these solitary waves are $c_{s,1} =
0.7721~ m/s$ and $c_{s,2} = 0.7796~ m/s$.  Their amplitudes are $A_1 =
0.0108~ m$ and $A_2 = 0.0120~ m$ respectively. (In this experiment the
depth $d=0.05~ m$.).  At $t = 18.3~ s$ these solitary waves achieved
their maximum values at $x_1 = 0.247~ m$ and $x_2 = 1.348~ m$
respectively.  Figures~\ref{fig:figure7} and \ref{fig:figure8} include
comparisons between the numerical solution and experimental data.  The
agreement between the numerical results and the experimental data is
impressive.  The agreement in the generated dispersive tails in
Figure~\ref{fig:figure8} is even more impressive. Such agreement
cannot be found in the case of head-on collisions of solitary waves of
Boussinesq type models, \cite{Dutykh2011e}, indicating that the
high-order nonlinear terms are important in studying even these
small-amplitude solutions. Finally, we mention that the maximum
amplitude of the solution observed in Figure \ref{fig:figure7}(c)
during the collision is smaller than the real amplitude, possibly,
because of a splash phenomenon that cannot be described by any model
(see also \cite{Dutykh2011e}).

\begin{figure}
  \bigskip
  \centering
  {\includegraphics[width=0.99\columnwidth]{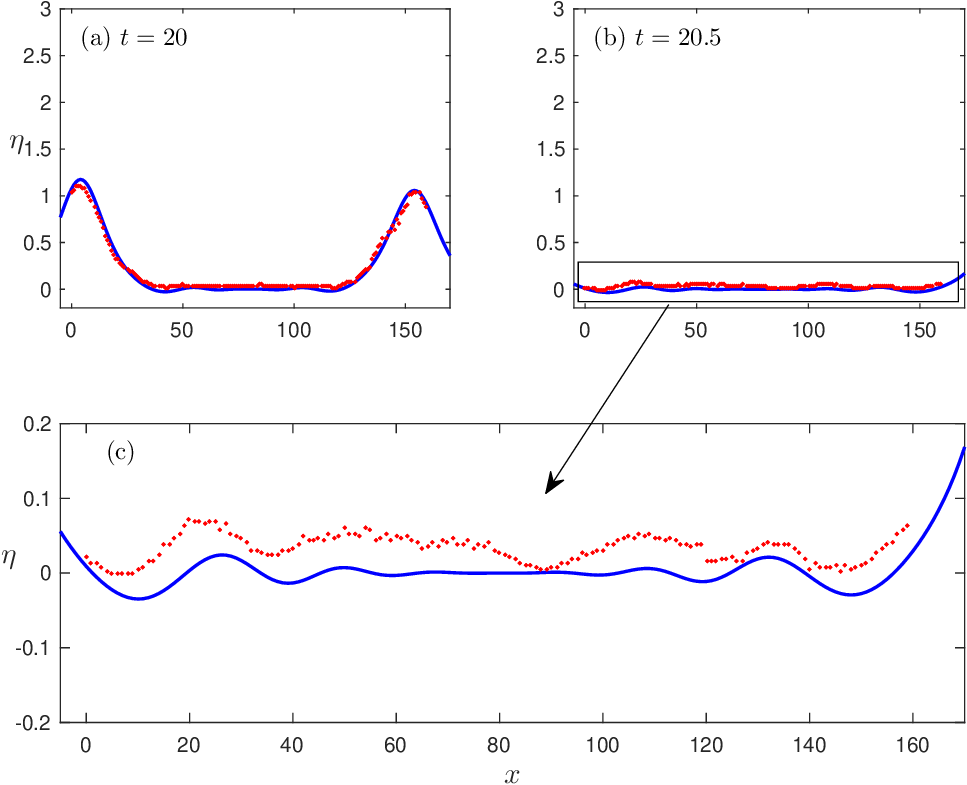}}
  \caption{\small\em (Cont'd) Comparison of the head-on collision of
    two solitary waves of the \textsc{Serre} system with experimental data.} 
  \label{fig:figure8}
\end{figure}

\begin{figure}
  \bigskip
  \centering
  {\includegraphics[width=0.99\columnwidth]{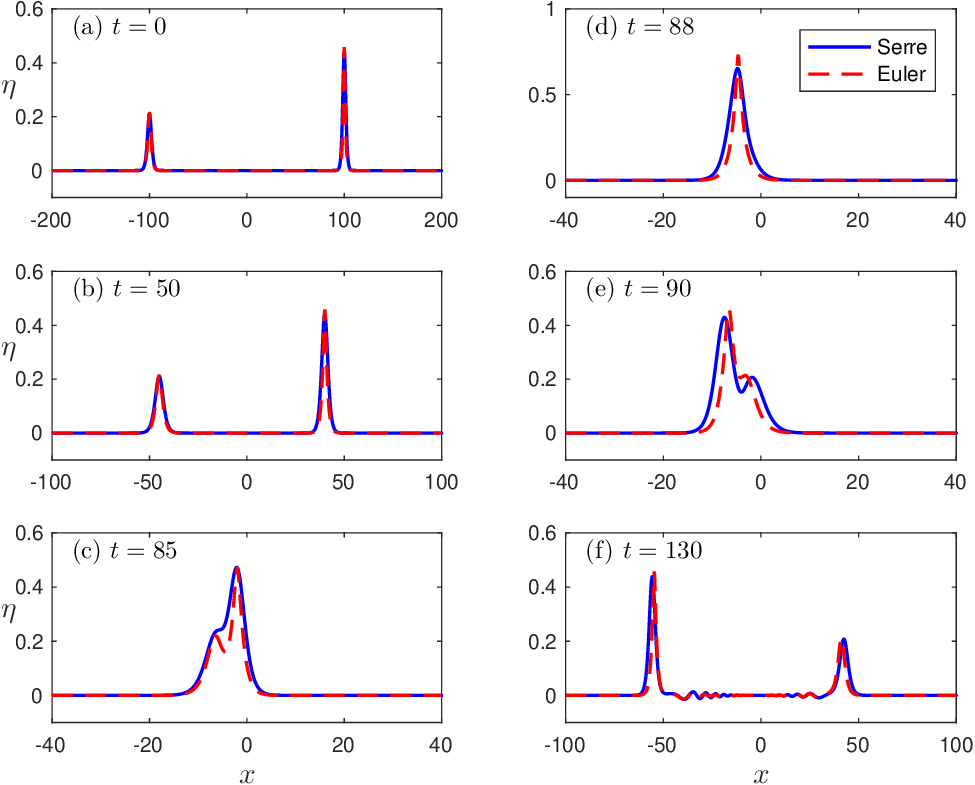}}
  \caption{\small\em Comparison between the head-on collisions of two
    solitary waves for the \textsc{Serre} and \textsc{Euler} systems.}
  \label{fig:figure9}
\end{figure}

We now compare a head-on collision of two unequal solitary waves via
numerical solutions of the \textsc{Serre} and \textsc{Euler} equations.  For both
models, we consider a right-traveling solitary wave with $c_s=1.1$ and
a left-traveling solitary wave with $c_s = 1.2$.  These solitary waves
are initially translated so that the maximum peak amplitudes are
achieved at $x=-100$ and $x=100$ respectively. Results from the
numerical simulations are included in Figure~\ref{fig:figure9}.  Both models show similar behavior, however the 
maximum amplitude observed during the collision using the \textsc{Euler}
equations is larger than in the \textsc{Serre} system. Also the interaction in
the \textsc{Euler} equations lasts longer and therefore a larger phase shift is
observed.  The leading waves of the dispersive tails are almost identical in the two models, but the
amplitude of the tails in the \textsc{Euler} system decay to zero more slowly
than the amplitude of the tails in the \textsc{Serre} system.  These numerical
simulations verify the ability of the \textsc{Serre} system to accurately model
head-on collisions of solitary waves.  They also show that the \textsc{Serre}
system is consistent with the \textsc{Euler} equations during and after the
head-on collision with almost identical solutions.

\section{Stability of traveling waves}\label{sec:stability}

The previous experiment of the head-on collision of two solitary waves
indicates that the solitary waves are robust. In this section, we
present the behavior of a solitary wave under small perturbations.  We
explore the effects of modifying some of the high-order terms of the
\textsc{Serre} system. We show that modifying one such term one can produce
{\em regularized} shock waves, as opposed to classical dispersive
shock waves.  Finally, we examine the stability of the cnoidal wave
solutions.

\subsection{Stability of solitary waves}\label{sec:pertsw}

We consider perturbations of the amplitude, perturbations of the
wavelength, and random-noise perturbations of the shape.  As we show
below, all of the solitary waves we tested were stable to all of the
perturbations we considered.

\begin{figure}
  \bigskip
  \centering
  {\includegraphics[width=0.99\columnwidth]{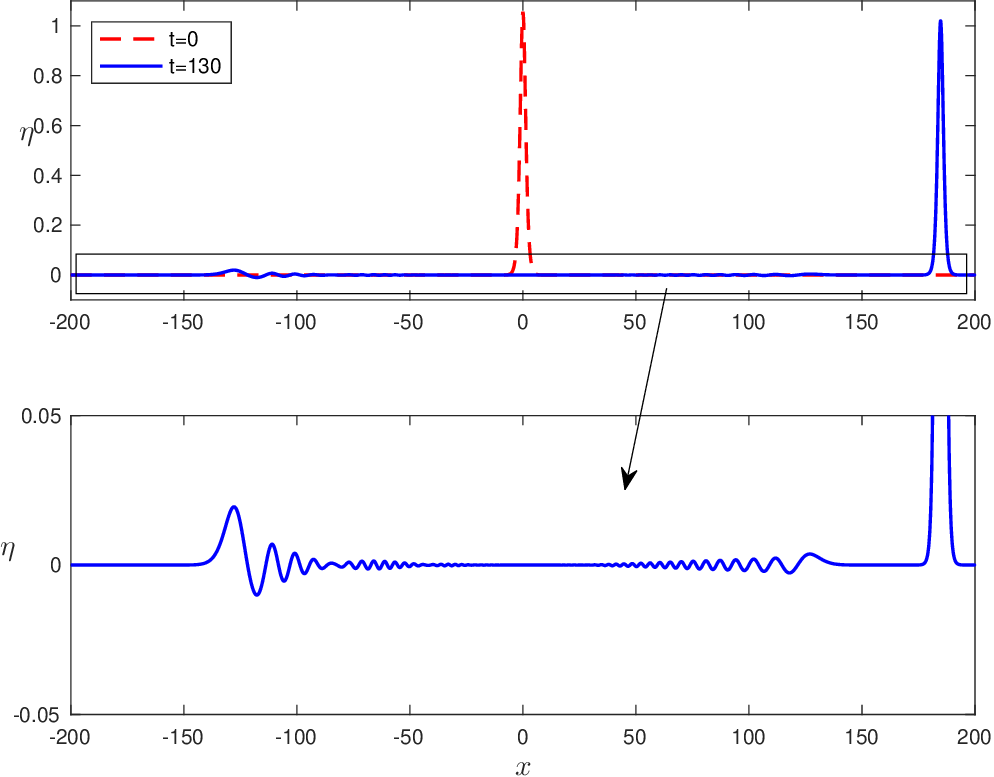}}
  \caption{\small\em The perturbed solution generated by the
    perturbation of the amplitude of the solitary wave with $c_s =
    1.4$, $A=0.96$, and perturbation parameter $p = 1.1$.} 
  \label{fig:figure11}
\end{figure}

We chose a solitary wave with speed $c_s=1.4$ and amplitude $A=0.96$
for all numerical simulations in this section.  We perturb the 
amplitude by multiplying the pulse by a parameter $p$ such that
\begin{equation}\label{eq:pertsolwave}
  h_p(x,0) = 1 + p\cdot a_1{\sech}^2(K_s\, x)\ ,  
\end{equation}
while keeping the velocity component of the solution unperturbed as in
(\ref{eq:solwave}).  When $p = 1.1$ the initial condition sheds a
small-amplitude dispersive tail and results in a new solitary wave
with amplitude $A = 1.02050$.  Figure~\ref{fig:figure11} presents the
initial condition and the resulting solution at $t=130$.  Similar
observations resulted in all cases we tested. 

We consider perturbations of the wavelength $K_s$ by taking the
initial condition for $h$ to be 
\begin{equation}\label{eq:pertsolwave2}
  h_p(x,0) = 1 +  a_1{\sech}^2(p\cdot K_s\, x)\ .
\end{equation}
The results in this case were very similar to the results we obtained
when we perturbed the amplitude of the solitary waves and so we don't
show the results here. Table~\ref{tab:pert1}  shows the amplitudes of
the solitary waves that result from various amplitude and wavelength
perturbations.

\begin{table}[ht]
\center
  \begin{tabular}{lll}
    \hline
    $p$ & Amplitude perturbation & Wavelength perturbation \\
    \hline
    $0.8$ & $0.83860$ & $1.02710$\\
    $0.9$ & $0.89936$ & $0.99225$\\
    $1.1$ & $1.02050$ & $0.93017$\\
    $1.2$ & $1.08087$ & $0.90260$\\
    \hline
  \end{tabular}
  \bigskip
  \caption{\small\em Amplitudes of the uniformly perturbed solitary waves.}\label{tab:pert1} 
\end{table}

\begin{figure}
  \bigskip
  \centering
  {\includegraphics[width=0.99\columnwidth]{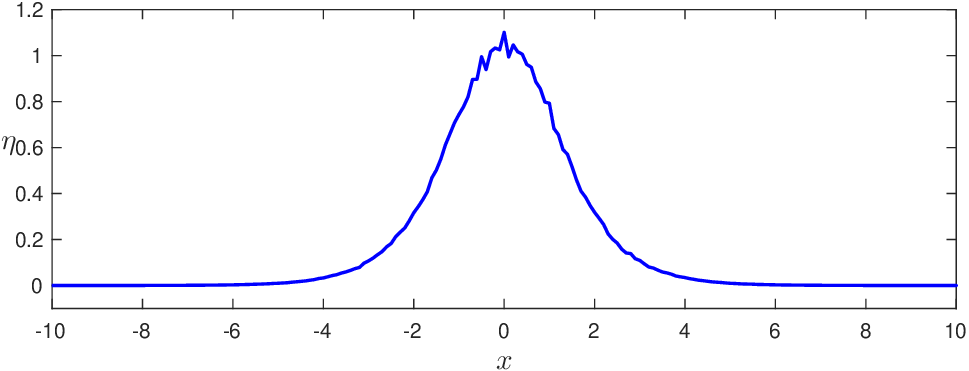}}
  \caption{\small\em The solitary wave perturbed by pseudo-random
    noise with $p = 0.2$.}
  \label{fig:figure12}
\end{figure}

\begin{figure}
  \bigskip
  \centering
  {\includegraphics[width=0.79\columnwidth]{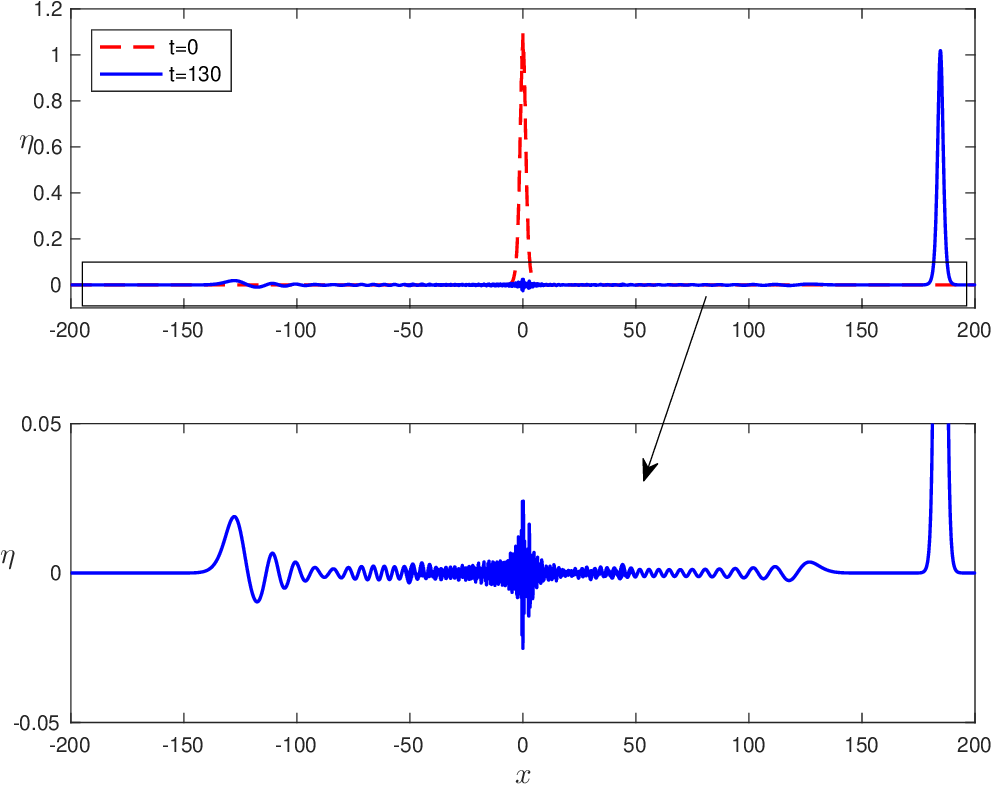}}
  \caption{\small\em Evolution of the initial condition shown in Figure~\ref{fig:figure12} with $p = 0.2$.}
  \label{fig:figure13}
\end{figure}

Similar results were obtained when non-uniform perturbations were
used.  In order to consider non-uniform perturbations, we used
pseudo-random noise distributed uniformly in $[0,1]$.  Denoting the
noise function by $N(x)$, the perturbed solitary wave is given by
\begin{equation}
h_p(x,t) = 1 + \bigl(1-p\ N(x)\bigr)\cdot a_1{\sech}^2(K_s\, x-c_st)\ ,
\end{equation}
where the parameter $p$ determines the magnitude of the
noise.  Figure~\ref{fig:figure12} shows the perturbed solitary wave
with $p = 0.2$. This type of perturbation is not only non-uniform, but
is also non-smooth. Nevertheless this initial condition is the
$L^2$-projection of the actual solution which ensures the required by
the FEM smoothness. Figure~\ref{fig:figure13} shows the evolution of
this perturbed solitary wave. The solution consists of a new solitary
wave and a small-amplitude dispersive tail.  It does not differ
qualitatively from the solution shown in Figure~\ref{fig:figure11}.
The values of the amplitudes of the emerging solitary waves for
various values of $p$ are presented in Table~\ref{tab:pert2}. These
results suggest that the solitary waves of the \textsc{Serre} system are
orbitally stable with respect to this class of perturbations.

\begin{table}[ht]
\center
  \begin{tabular}{ll}
    \hline
    $p$ & Amplitude \\
    \hline
    $-0.2$ & $0.90110$ \\
    $-0.1$ & $0.93055$ \\
  $0.1$ & $0.98942$ \\
  $0.2$ & $1.01883$ \\
  $0.5$ & $1.10694$ \\
  \hline
  \end{tabular}
  \bigskip
  \caption{\small\em Amplitudes of the randomly perturbed solitary waves.}
  \label{tab:pert2} 
\end{table}

\subsection{Persistence of the solitary waves}\label{sec:persist}

One other aspect related to the stability of the solitary waves is
their ability to persist when some of the high-order terms in the PDE
are perturbed. In this section, we examine if a \textsc{Serre} solitary wave
retains its shape when some of the terms of the \textsc{Serre} system are
perturbed. Introducing the parameters $\alpha$, $\beta$, and $\gamma$,
we rewrite the \textsc{Serre} equations in the following form
\begin{equation}
\begin{array}{l}
\eta_t + u_x + (\eta u)_x = 0~ , \\
u_t + \eta_x + uu_x - \frac{1}{3h}[h^3(\alpha\,u_{xt} + \beta uu_{xx} - \gamma (u_x)^2)]_x = 0~ .
\end{array}\label{eq:nserre}
\end{equation}
The unperturbed \textsc{Serre} equations correspond to
$\alpha=\beta=\gamma=1$. We first study the persistence of the solitary
waves when the system is perturbed by perturbing the parameters
$\alpha$, $\beta$, $\gamma$ and considering a solitary wave of the
unperturbed system as an initial condition.  In this section we use
the solitary wave (\ref{eq:solwave}) 
with $c_s=1.4$ as an initial condition. If $\alpha = 0.9$, or if
$\beta=0.9$, or if $\gamma=0.9$ the solitary wave evolves in a manner
similar to the amplitude perturbations in Section~\ref{sec:pertsw}.
The new solitary waves are very similar to the unperturbed solitary
wave.  This further indicates that the solitary waves of the \textsc{Serre}
system are stable.  

\begin{figure}
  \bigskip
  \centering
  {\includegraphics[width=0.99\columnwidth]{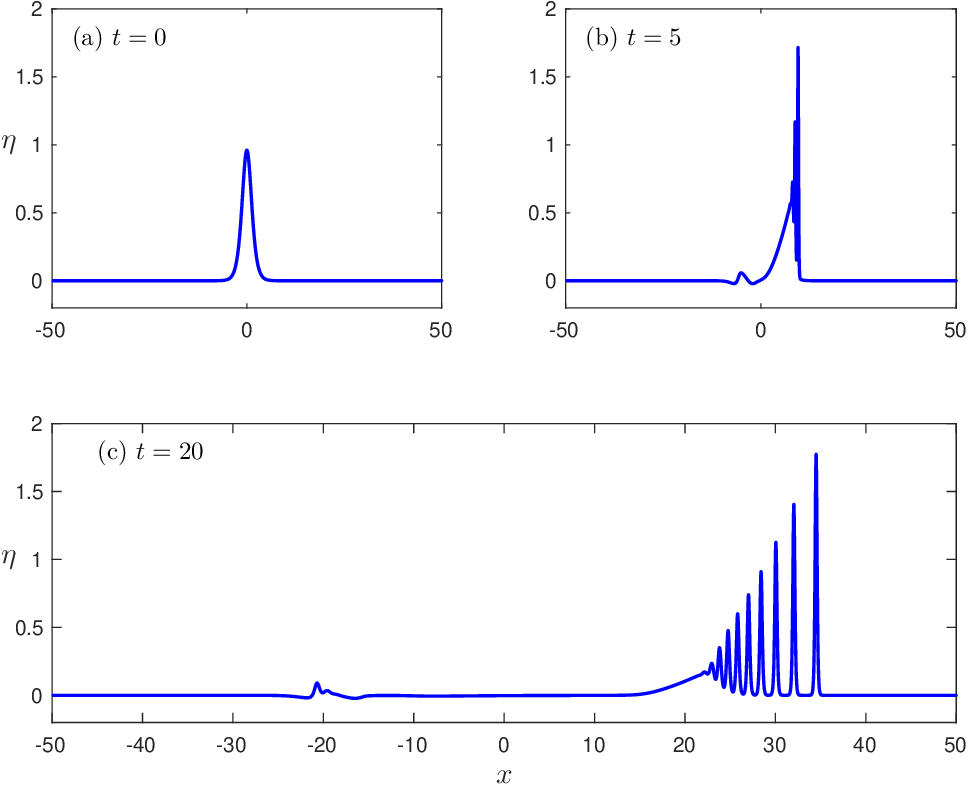}}
  \caption{\small\em The evolution of a solitary wave of a perturbed system ($\alpha = \beta = \gamma = 0.01$).}
  \label{fig:figure15}
\end{figure}

More interesting phenomena is observed when the solitary waves are
used as initial conditions to systems with small values of the
parameters $\alpha$, $\beta$ and $\gamma$.  When all the three
parameters are very small, the solutions tend to break into dispersive
shock waves or other forms of undular bores.  In the first numerical
simulation, we consider $\alpha=\beta=\gamma=0.01$.  This is similar
to the case of the small dispersion limit where the weakly nonlinear 
terms are dominant. Figure~\ref{fig:figure15} demonstrates that the
solution becomes a dispersive shock. This phenomenon has been
previously observed in dispersive systems, \cf
\cite{Dutykh2011e,El2006,Lax1983}.  Unexpectedly, taking the
parameters $\beta$ and $\gamma$ to be very small, \ie $\beta = \gamma
= 0.001$ and keeping the parameter $\alpha = 1$, the solitary wave
persists and evolves into a new solitary wave which is similar to the
unperturbed solitary wave qualitatively similar to those presinted in Section~\ref{sec:pertsw}.
This persistence is remarkable because the solitary wave remains
almost the same even if two of the most important terms have been
almost eliminated.  If $\alpha=\beta=1$ and $\gamma=0.001$ or if
$\alpha=\gamma=1$ and $\beta=0.001$ the behavior is similar.

\begin{figure}
  \bigskip
  \centering
  {\includegraphics[width=0.99\columnwidth]{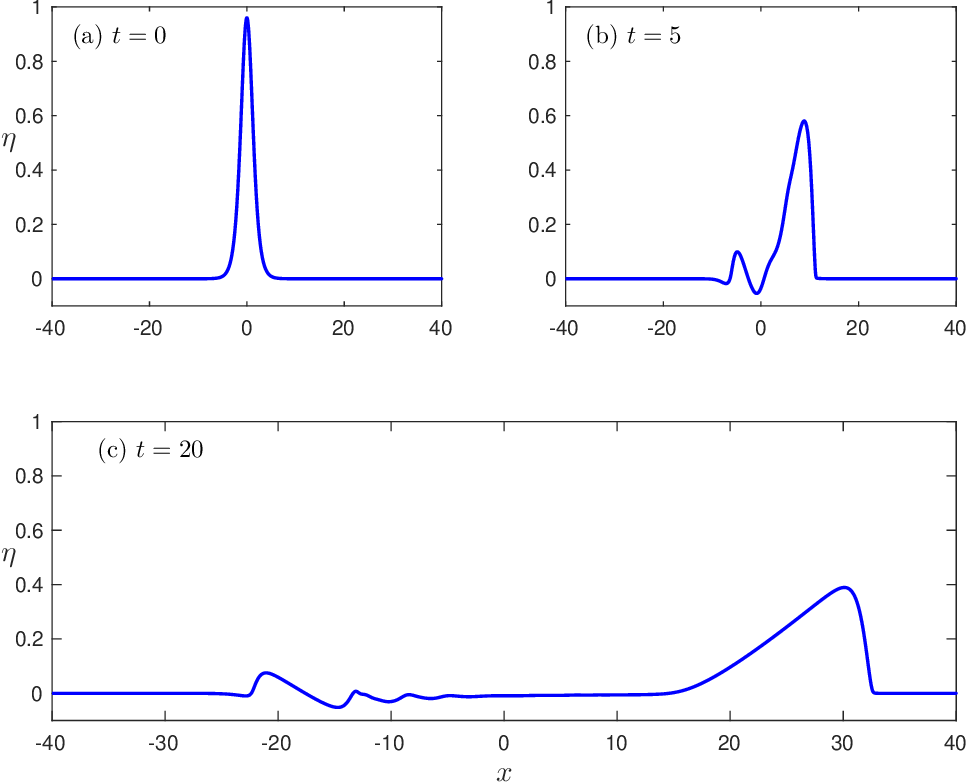}}
  \caption{\small\em The evolution of a solitary wave of a perturbed system and the generation of a regularized shock ($\alpha = 0.001$, $\beta = \gamma = 1$).}
  \label{fig:figure17}
\end{figure}

The behavior changes dramatically if large perturbations $\alpha$ are
considered.  The results from the simulation with $\alpha=0.001$ and
$\beta=\gamma=1$ is shown in Figure~\ref{fig:figure17}.  In this case,
the initial condition breaks into different waves but instead of
forming a dispersive shock wave, it forms a new kind of regularized
shock wave. This suggests a new breaking mechanism by the elimination
of the $u_{xt}$ term. Similar dissipative behavior has been observed
in nonlinear KdV-type equations where some high-order nonlinear terms
introduce dissipation to the system~\cite{Brenier2000}. 

\subsection{Stability of Cnoidal waves}\label{sec:stabcn}

We follow the work of Carter \& Cienfuegos~\cite{Carter2011} in order
to study the linear stability of the solutions given in
(\ref{eq:cnwave}).  We enter a coordinate
frame moving with the speed of the solutions by defining $\chi=x-c_c\
t$ and $\tau=t$. In this moving frame, the \textsc{Serre} equations are given
by
\begin{equation}
\begin{array}{l}
h_\tau - c_c h_\chi + \big{(}h u\big{)}_\chi = 0~ , \\
u_\tau - c_c u_\chi + uu_\chi + h_\chi-\frac{1}{3h}\Big{(}h^3\big{(}u_{\chi\tau} - c_c u_{\chi\chi} + uu_{\chi\chi} - (u_\chi)^2\big{)}\Big{)}_\chi = 0~ ,
 \end{array}\label{serremoving}
\end{equation}
and the solution given in (\ref{eq:cnwave}) simplifies to the following time-independent solution
\begin{equation}
\begin{array}{l}
h = h_0(\chi) = a_0 + a_1\mbox{dn}^2\big{(}K_c\chi,k\big{)}~ , \\
u = u_0(\chi) = c_c\ \Big{(}1-\frac{h_0}{h(\chi)}\Big{)}~ .
\end{array}\label{dnsolnstationary}
\end{equation}

We consider perturbed solutions of the form

\begin{equation}
\begin{array}{l}
h_{\mbox{pert}}(\chi,\tau) = h_0(\chi) + \mu h_1(\chi,\tau) + \O(\mu^2)~,\\
u_{\mbox{pert}}(\chi,\tau) = u_0(\chi) + \mu u_1(\chi,\tau) + \O(\mu^2)~ ,
 \end{array}\label{pertsoln}
\end{equation}
where $h_1$ and $u_1$ are real-valued functions and $\mu$ is a
small real parameter. Substituting (\ref{pertsoln}) into
(\ref{serremoving}) and linearizing leads to a pair of coupled, linear
partial differential equations that are constant coefficient in
$\tau$. Without loss of generality, assume 
\begin{equation}
\begin{array}{l}
h_1(\chi,\tau) = H(\chi)\ue^{\Omega \tau} + \mbox{c.c.}~ , \\
u_1(\chi,\tau) = U(\chi)\ue^{\Omega \tau} + \mbox{c.c.}~ ,
\end{array}\label{pertform}
\end{equation}
where $H(\chi)$ and $U(\chi)$ are complex-valued functions, $\Omega$
is a complex constant, and $\mbox{c.c.}$ denotes complex conjugate. If
$\Omega$ has a positive real part, \ie if $\Re(\Omega) > 0$, then the
perturbations $h_1$ and $u_1$ grow exponentially in $\tau$ and the
solution is said to be unstable. 

Substituting (\ref{pertform}) into the linearized PDEs gives
\begin{equation}\label{evprob}
  \mathcal{L}\left( \begin{array}{c} 
  H \\ U \end{array}\right) = \Omega\hspace{0.1cm} \mathcal{M}\left( \begin{array}{c} H \\ U \end{array} \right)~ ,
\end{equation}
where $\mathcal{L}$ and $\mathcal{M}$ are the linear differential
operators defined by 
\begin{equation}\label{LM}
\begin{array}{l}
\mathcal{L} = 
\left( \begin{array}{cc} 
  -u_0^{\prime}+(c_c-u_0)\partial_\chi & -\eta_0^{\prime}-\eta_0\partial_\chi \\ 
 \mathcal{L}_{21} & \mathcal{L}_{22} 
 \end{array}\right)~ ,\\
\mathcal{M} = 
 \left( \begin{array}{cc} 1 & 0 \\ 
  0 & 1 - \eta_0\eta_0^{\prime}\partial_\chi - 
  \frac{1}{3}\eta_0^2\partial_{\chi\chi} 
\end{array}\right)~ ,
\end{array}
\end{equation}
where prime represents derivative with respect to $\chi$ and
\begin{eqnarray}
 \mathcal{L}_{21} &= & -\eta_0^{\prime}(u_0^{\prime})^2-c_c\eta_0^{\prime}u_0^{\prime\prime}-\frac{2}{3}c_c \eta_0u_0^{\prime\prime\prime}+\eta_0^{\prime}u_0u_0^{\prime\prime}-\frac{2}{3}\eta_0u_0^{\prime}u_0^{\prime\prime}+ \nonumber \\ 
 & & \frac{2}{3}\eta_0u_0u_0^{\prime\prime\prime}+ \big{(}\eta_0u_0u_0^{\prime\prime}-g-\eta_0(u_0^{\prime})^2-c_c\eta_0u_0^{\prime\prime}\big{)}\partial_\chi~ , \\
  \mathcal{L}_{22} & = & -u_0^{\prime}+\eta_0\eta_0^{\prime}u_0^{\prime\prime}+\frac{1}{3}\eta_0^2u_0^{\prime\prime\prime}+\big{(}c_c-u_0-2\eta_0\eta_0^{\prime}u_0^{\prime}-\frac{1}{3}\eta_0^2u_0^{\prime\prime}\big{)}\partial_\chi+\nonumber \\
  & & \big{(}\eta_0\eta_0^{\prime}u_0-c_c\eta_0\eta_0^{\prime}-\frac{1}{3}\eta_0^2u_0^{\prime}\big{)}\partial_{\chi\chi}+\big{(}\frac{1}{3}\eta_0^2u_0-\frac{1}{3}c_c \eta_0^2\big{)}\partial_{\chi\chi\chi}~ .
\end{eqnarray}

The Fourier--Floquet--Hill method described in Deconinck \&
Kutz~\cite{Deconinck2006} is then used to solve the differential
eigenvalue problem given in (\ref{evprob}). This method establishes
that all bounded solutions of (\ref{evprob}) have the form 
\begin{equation}\label{efxnform}
  \left( \begin{array}{c} H \\ U \end{array}\right) = 
  \ue^{\ui\rho\chi} \left(\begin{array}{c} H^P \\ U^P \end{array}\right)~ ,
\end{equation}
where $H^P$ and $U^P$ are periodic in $\chi$ with period $2K/K_c$ and
$\rho \in \bigl[-\pi K_c/(4K), \pi K_c/(4K)\bigr]$. 

\begin{figure}
  \bigskip
  \begin{center}
  \includegraphics[width=0.79\columnwidth]{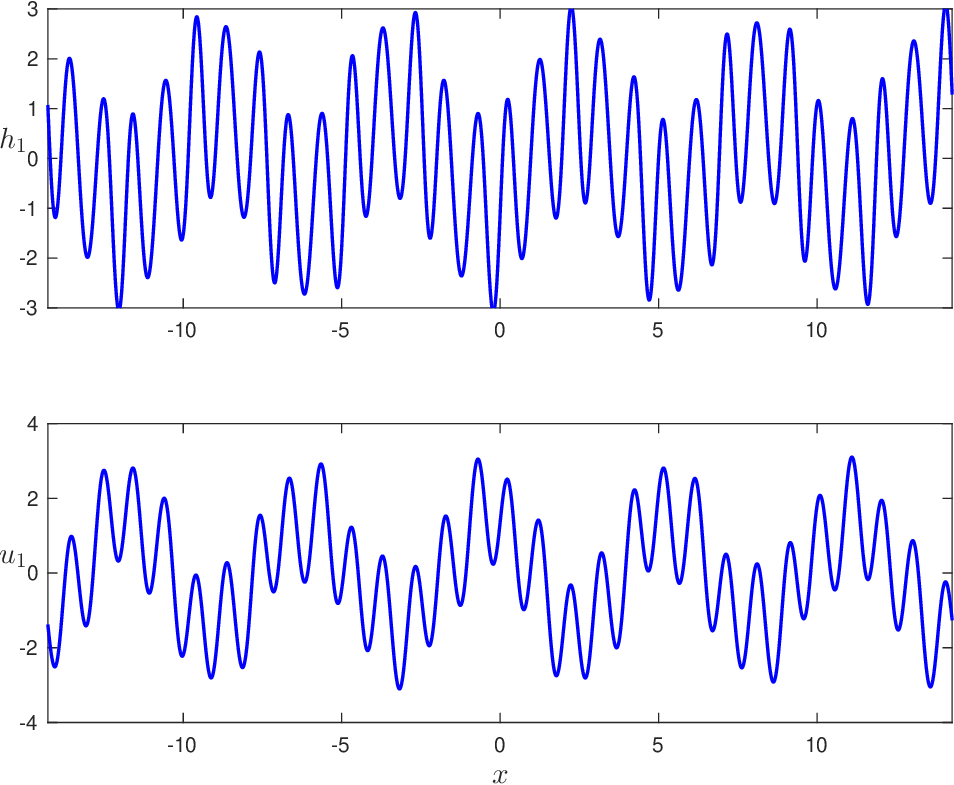}
  \caption{\small\em Unstable perturbations corresponding to the solution given in (\ref{eq:cnwave}) with $a_0 = 0.3$, $a_1 = 0.2$ and $k = 0.75$.}
  \label{fig:unstablemode}
  \end{center}
\end{figure}

\begin{figure}
  \bigskip
  \begin{center}
  \includegraphics[width=0.99\columnwidth]{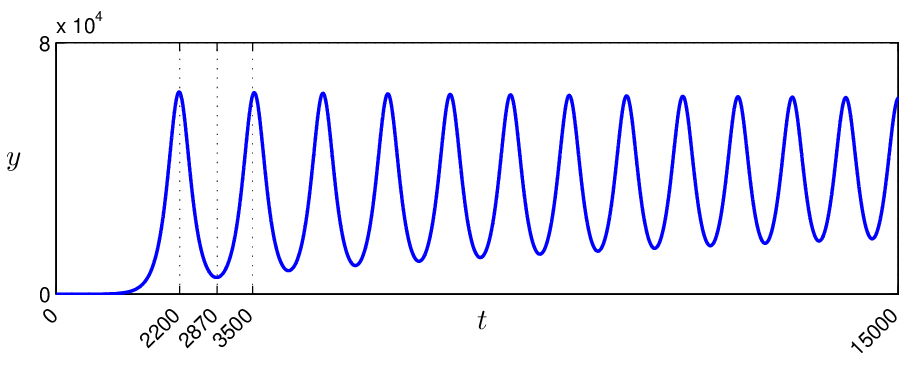}
  \caption{\small\em The magnitude of the first Fourier mode of the perturbation versus time.}
  \label{fig:figure19}
  \end{center}
\end{figure}

\begin{figure}
  \bigskip
  \begin{center}
  \includegraphics[width=0.99\columnwidth]{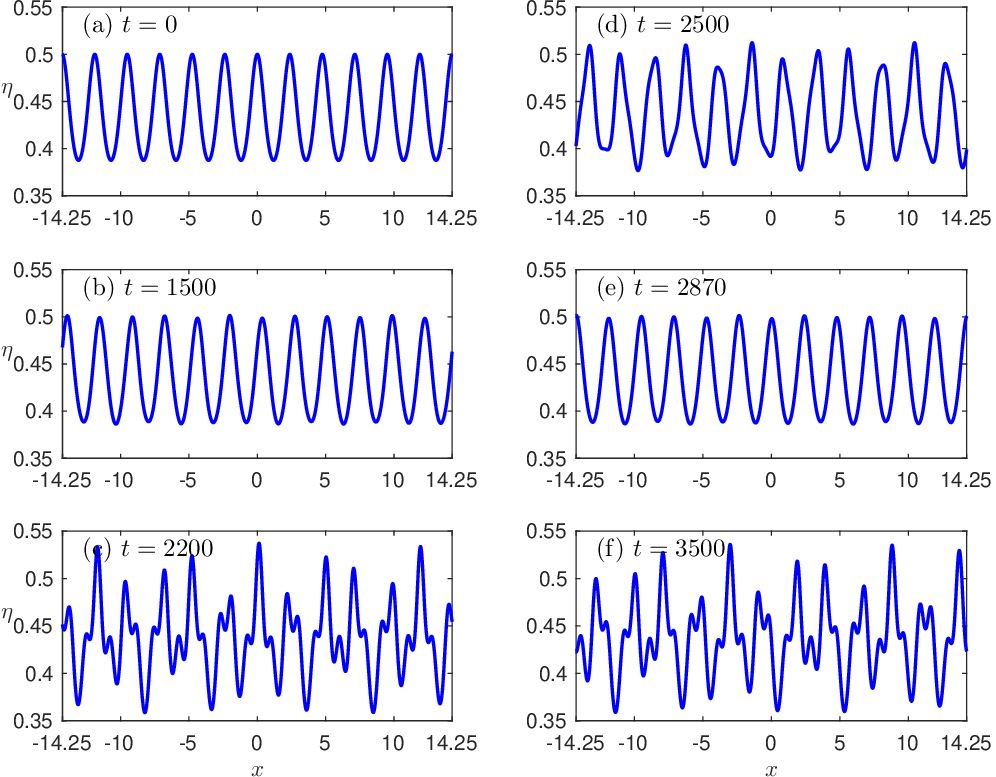}
  \caption{\small\em The periodic instability of the perturbed cnoidal wave.}
  \label{fig:figure20}
  \end{center}
\end{figure}

Using this method, Carter \& Cienfuegos established that solutions of
the form given in (\ref{dnsolnstationary}) with sufficiently small
amplitude and steepness are spectrally stable and solutions with
sufficiently large amplitude or steepness are spectrally unstable. For
example, the solution with $a_0 = 0.3$, $a_1 = 0.2$ and $k = 0.75$ is
unstable with respect to the perturbation shown in
Figure~\ref{fig:unstablemode}. The period of this perturbation is
twelve times the period of the exact solution ($\rho = 1/12$). The
theory establishes that the magnitude of this perturbation will grow
like $\ue^{0.00569 t}$. We corroborated this result by using the
following perturbation-seeded solution as the initial condition in our
\textsc{Serre} solver
\begin{equation}\label{pertsoln222}
\begin{array}{l}
h_{pert}(x,0) = h(x,0) + 10^{-7}h_1(x)~ ,\\
u_{pert}(x,0) = u(x,0) + 10^{-7}u_1(x)~ .
\end{array}
\end{equation}
Here $\big{(}h(x,0), u(x,0)\big{)}$ is the solution given in equations
(\ref{dnsolnstationary}) with $a_0 = 0.3$, $a_1 = 0.2$ and $k = 0.75$
and $\big{(}h_1(x), u_1(x)\big{)}$ is the perturbation shown in
Figure~\ref{fig:unstablemode}. Figure~\ref{fig:figure19} contains a
plot of the magnitude of the first Fourier mode of the solution versus
$t$. This mode initially (up to $t = 1500$) grows exponentially with a
rate of $0.00577$, very close to the rate predicted by the linear
theory.  However after more time, the solution returns to a state
close to the initial one.  The first portion of this recurrence
phenomenon is depicted in Figure~\ref{fig:figure20}.  Note that the
solution at $t=2870$ has nearly returned to its initial state. Similar
behavior has been observed to other shallow water models by Ruban
\cite{Ruban2012} and it is referred to as the Fermi-Pasta-Ulam
recurrence.

\section{Dispersive shock waves}\label{sec:dswinter}

A simple DSW traveling to the right can be generated using the
Riemann initial data, \cf \cite{El2006},
\begin{equation}\label{eq:DSW-init}
  h(x,0) = \left\{
  \begin{array}{ll}
    h^{-}, & \mbox{ for } x<0\\
    h^{+}, & \mbox{ for } x>0
  \end{array}
  \right. ,
\quad 
u(x,0) = \left\{
  \begin{array}{ll}
    u^{-}, & \mbox{ for } x<0\\
    u^{+}, & \mbox{ for } x>0
  \end{array}
  \right.~,
\end{equation}
with the compatibility condition (Riemann invariant) 
\begin{equation}\label{eq:u-}
  \frac{u^{-}}{2} - \sqrt{h^{-}} = \frac{u^{+}}{2} - \sqrt{h^{+}}~.
\end{equation}

\begin{figure}
  \bigskip
  \begin{center}
  \includegraphics[width=0.99\columnwidth]{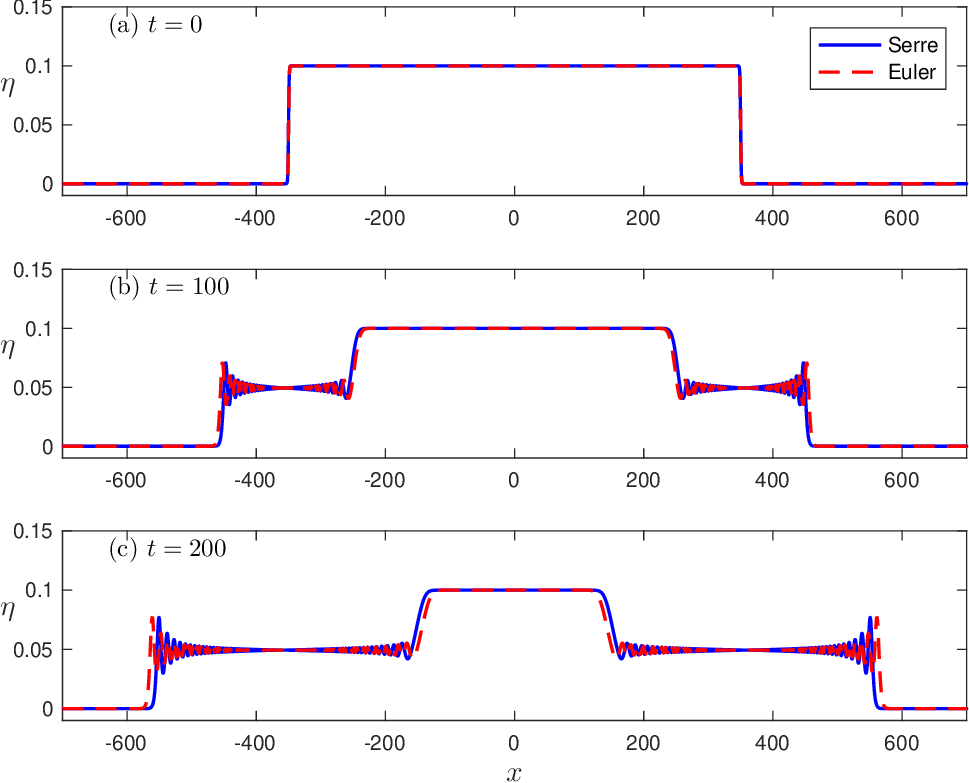}
  \caption{\small\em The dam break problem.}
  \label{fig:figure22}
  \end{center}
\end{figure}

DSWs can also be generated during the dam-break problem
simulation. In this case, the initial data for $h(x,0)$ are the same
as in (\ref{eq:DSW-init}), but there is no flow at $t = 0$, \ie,
$u(x,0) = 0$. As shown in \cite{El2006}, this generates two
counter-propagating DSWs, one on each side of the ``\emph{dam}'', and
two rarefaction waves that travel toward the center. We consider the
initial condition for $h$ to be a smooth step function that decays to
zero as $|x| \to \infty$.  Specifically, we choose
\begin{equation}\label{eq:h0}
  \eta(x,0) \ =\ \frac{1}{2}\eta_0 \left[1 + \tanh \left(\frac{x_0 - |x|}{2}\right)\right],
\end{equation}
where $\eta_0 = 0.1$, $x_0 = 350$, and $u(x,0) = 0$.  A plot of this
initial condition is included in Figure \ref{fig:figure22}.  Both the
\textsc{Euler} and \textsc{Serre} equations generate two counter propagating DSWs
and two rarefaction waves.  Figure~\ref{fig:figure22} demonstrates
that the amplitude of the leading wave for both solutions is
almost the same.  For example, the amplitude of the \textsc{Euler} leading wave
at $t = 200$ is $A = 0.06372$ while the amplitude of the \textsc{Serre}
leading wave at $t=200$ is $A = 0.06356$. Although the leading waves
have almost the same amplitudes, the phase speeds are slightly
different.  The difference in phase speeds is demonstrated in
Figure~\ref{fig:figure22}. 

\begin{figure}
  \bigskip
  \begin{center}
  \includegraphics[width=0.99\columnwidth]{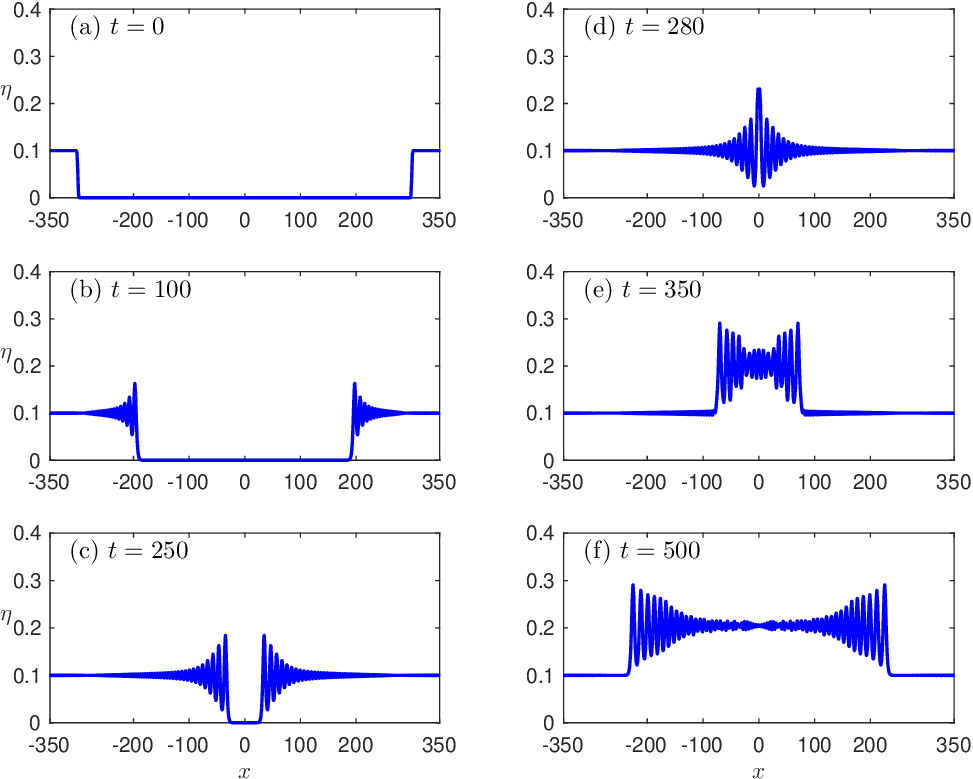}
  \caption{\small\em Head-on collision of two simple DSWs.}
  \label{fig:figure23}
  \end{center}
\end{figure}

After verifying that the \textsc{Serre} system has dispersive shock waves
that are comparable with the full \textsc{Euler} equations, we examine the
interactions of simple DSWs starting with the head-on collision.  For
the head-on collision we again consider two initial waveforms similar
to (\ref{eq:DSW-init}) but translated as is shown  in Figure
\ref{fig:figure23}(\textit{a}).  These step functions generate two
counter-propagating waves that begin to interact at approximately
$t=27$. The collision is inelastic.  After the collision there are two
DSWs propagating in different directions on the trailing edge of the
DSWs.

\begin{figure}
  \bigskip
  \begin{center}
  \includegraphics[width=0.99\columnwidth]{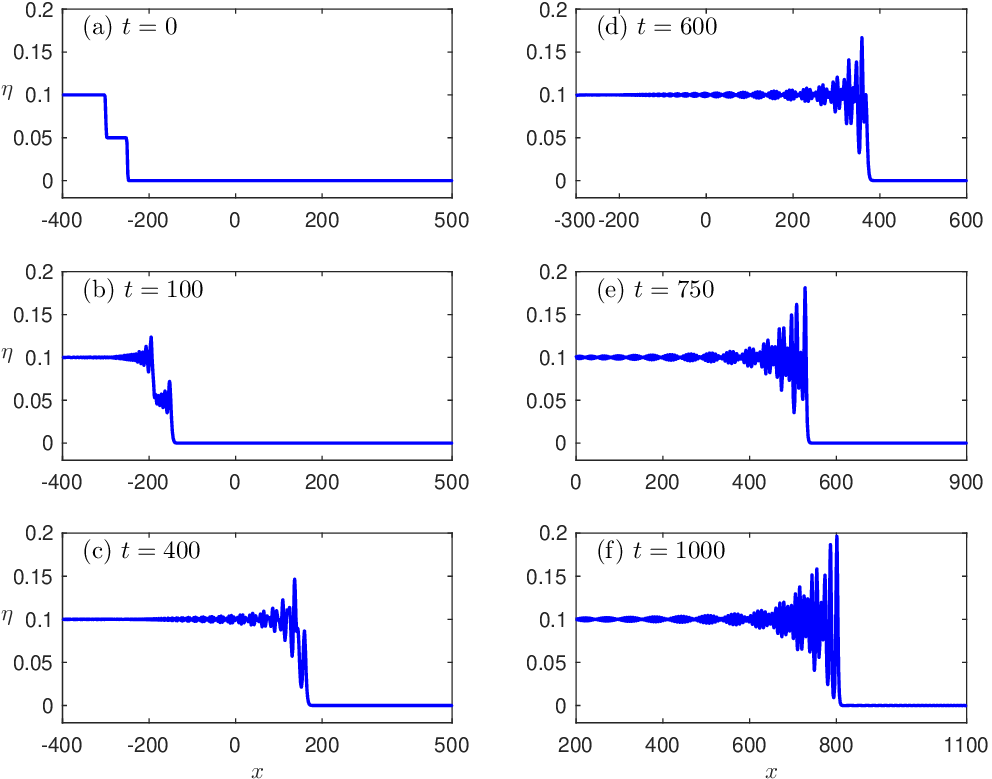}
  \caption{\small\em Overtaking interaction of two DSWs.}
  \label{fig:figure24}
  \end{center}
\end{figure}

We now consider overtaking collisions of DSWs. For this situation we
consider double-step initial conditions as is shown in
Figure~\ref{fig:figure24}. The first step has amplitude $0.1$ while
the shorter step has amplitude $0.05$.  This initial condition
generate two DSWs that propagate to the right. Because shorter
DSWs propagate with smaller phase speeds than taller DSWs, the taller
DSW approaches the shorter one and they interact.  The interaction is
so strong that the symmetry of the leading wave of both DSWs is
destroyed.  The two waves appear to merge and propagate as one 
single-phase DSW. Similar behavior has been observed in NLS-type and
KdV-type equations \cite{Hoefer2007,Ablowitz2013}.  Finally, we
mention that the solutions shown in Figures~\ref{fig:figure23} and
\ref{fig:figure24} are magnifications of the actual solutions. The
rest of the solution, not shown in these figures, consists of
dispersive rarefaction waves that we do not study in this paper.  For
more information see \cite{Mitsotakis2014}. 

\section*{Acknowledgments}

D.~\textsc{Mitsotakis} thanks Professor Boaz \textsc{Ilan} for suggestions, comments, and stimulating discussions related to dispersive waves.  The authors
acknowledge the invaluable help of Professor Paul \textsc{Milewski} for
discussions related to the numerical schemes for the \textsc{Euler} equations
and Professor Didier \textsc{Clamond} for discussions on pseudo-spectral
methods. J.~\textsc{Carter} was supported by the National Science
Foundation under grant number DMS-1107476. D.~\textsc{Mitsotakis} was supported by the Marsden Fund administered by the Royal Society of New Zealand.


\addcontentsline{toc}{section}{References}
\bibliographystyle{abbrv}
\bibliography{biblio}

\end{document}